\documentclass{ws-procs10x7}

\begin{document}

\title{Electroweak Symmetry Breaking Circa 2005}

\author{S. Dawson}

\address{Physics Department, Brookhaven National Laboratory, Upton, NY, 11973,
USA\\E-mail: dawson@bnl.gov}

\twocolumn[\maketitle\abstract{Recent progress in both the
experimental and theoretical explorations of electroweak symmetry breaking is 
surveyed.}]

\section{Introduction}%1
 
\begin{figure}
\epsfxsize120pt
\figurebox{120pt}{160pt}{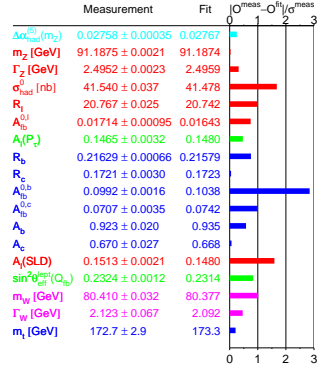}
\caption{Precision electroweak measurements and the best theoretical 
fit to the Standard Model as of September, 2005.  Also shown is the
deviation of the fit for each measurement from the value predicted using
the parameters of the central value of 
the fit[1].
%\cite{lepewwg}
}
\label{fig:pull}
\end{figure}

Particle physicists have a Standard Model of electroweak interactions
 which describes
a large number of measurements extraordinarily
well at energies on the few hundred $GeV$ scale.
  In fact, we have become extremely blas\'{e} about
tables such as that of Fig.~\ref{fig:pull},\cite{lepewwg} 
which shows an impressive agreement between
experiment and theory.  Virtual probes, using the sensitivity
of rare decays to high scale physics, are also in good agreement
with the predictions of the Standard Model.
This agreement, however, assumes the existence of a 
light, scalar Higgs boson, without which the theory is incomplete.
There has thus been an intense experimental effort at the Tevatron
aimed at discovering either the Standard Model Higgs boson or one
of the Higgs bosons associated with the minimal supersymmetric
model (MSSM).

\begin{figure}%1
\vskip -.5in
\epsfxsize120pt
\figurebox{120pt}{160pt}{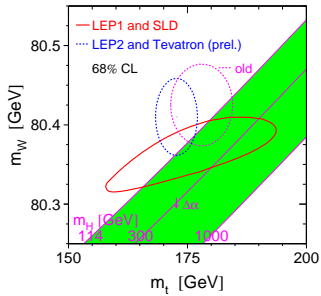}
\caption{The relationship between $M_W$ and $M_t$ in the Standard Model.
 The curve
labelled ``old'' does not include the Summer, 2005 updates on the
$W$ boson mass and width from LEP-2 and the new top quark mass
from the Tevatron[1].}
\label{fig:higglim}
\end{figure}

In the Standard Model, using $G_F$, $\alpha$, and  $M_Z$ 
as inputs, along with the 
fermion masses, the $W$ mass is a predicted quantity.  The
comparison between the prediction and the measured value can not only be used 
to check the consistency of the theory, but also to infer limits on possible 
extentions of the Standard Model.
The relationship between $M_W$ and $M_t$ is shown in Fig.~ \ref{fig:higglim}.
The curve labelled ``old'' does not include the new values
(as of Summer, 2005),  for the $W$ mass
and width from LEP-2 and the new mass of the top quark from the Tevatron.
(These new values are reflected in Fig.~ \ref{fig:pull}.)    

The measurements of Fig.~ \ref{fig:pull} can  be used to extract limits
on the mass of
a Standard Model Higgs boson.  The limit on the Higgs boson mass 
depends quadratically on the top quark mass and logarithmically on the
Higgs boson mass, making the limit
exquisitely sensitive to the top quark mass.  The limit is also quite
sensitive to which pieces of data are included in the analysis.  The fit
of Fig.~ \ref{fig:higglim} includes only the high energy data and so
does not include results from NuTeV or atomic parity violation.

The 
precision electroweak measurements of Fig.~\ref{fig:pull} give a $95\%$
confidence level upper limit on the value of the Higgs boson mass
of,\cite{lepewwg}
\begin{equation}
M_H < 186~GeV.
\end{equation}
If the LEP-2 direct search limit of $M_H > 114~GeV$ is included, the limit
increases to 
\begin{equation}
M_H < 219~GeV.
\end{equation}

\begin{figure}%1
\epsfxsize220pt
\figurebox{220pt}{220pt}{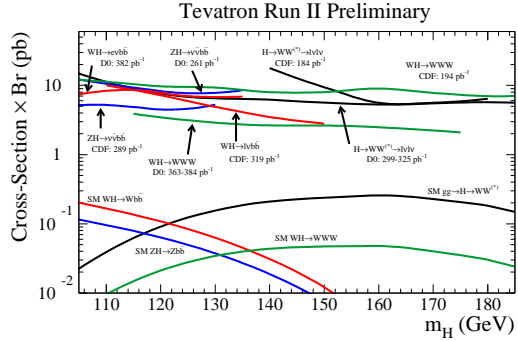}
\caption{CDF and D0 limits on the production cross section times
the branching ratios for various Higgs boson production channels
 as a function of the Higgs boson mass, along with the
Standard Model expectations for each channel[2].}
\label{fig:tevhig}
\end{figure}

Both CDF and D0 have presented experimental
limits on the production rate for a Standard Model Higgs
boson, which are shown in Fig.~ \ref{fig:tevhig}.\cite{cdfd0} For most
channels, the limits are still several orders of magnitude 
away from the predicted cross sections in
the Standard Model. With an integrated
luminosity of $4~fb^{-1}$ ($8~fb^{-1}$), the $95\%$ exclusion
limit will increase to $M_H>130~GeV$ ($M_H>135~GeV$).  A much more
optimistic viewpoint is to note that with $4~fb^{-1}$ there is a $35\%$
chance that the Tevatron will find $3\sigma$ evidence for a Higgs boson
with a mass up to $M_H=130~GeV$.

Despite the impressive agreement between the precision
electroweak data  and the theoretical 
predictions of the Standard Model
with a light Higgs boson, theorists have been busy inventing
new models where mechanisms other than a light Higgs boson are responsible
for the electroweak symmetry breaking.  We begin in Section \ref{needhiggs}
by reviewing the theoretical arguments for the existence of a Higgs boson
and continue in Section 3 to discuss the reasons why a light Higgs boson is
unattractive to many theorists.  In the following sections, we review a
sampling of models of electroweak symmetry breaking.

\section{Who needs a Higgs Boson?}
\label{needhiggs}

The Standard Model requires a Higgs boson for consistency with precision
electroweak data, as is clear from Fig.~ \ref{fig:higglim}. 
The Standard Model Higgs boson also serves two additional
critical functions.

The first is to generate gauge invariant masses for the fermions.  Since
left- ($\psi_L$) and right- ($\psi_R$) handed fermions transform differently
under the chiral $SU(2)_L \times U(1)_Y$ gauge groups, a mass term of the 
form
\begin{equation}
L_{mass}\sim m_f \biggl({\overline \psi_L}\psi_R+
{\overline \psi_R}\psi_L\biggr)
\end{equation}
is forbidden by the gauge symmetry.  A Higgs doublet, $\Phi$, with
a vacuum expectation value, $v$, generates a mass term of the required form,
\begin{equation}
L_{mass}\sim {m_f\over v} \biggl({\overline \psi_L}\Phi \psi_R +
{\overline \psi_R}\Phi^\dagger\psi_L\biggr).
\end{equation}

\begin{figure}%1
\epsfxsize240pt
\vskip -.6in
\figurebox{240pt}{240pt}{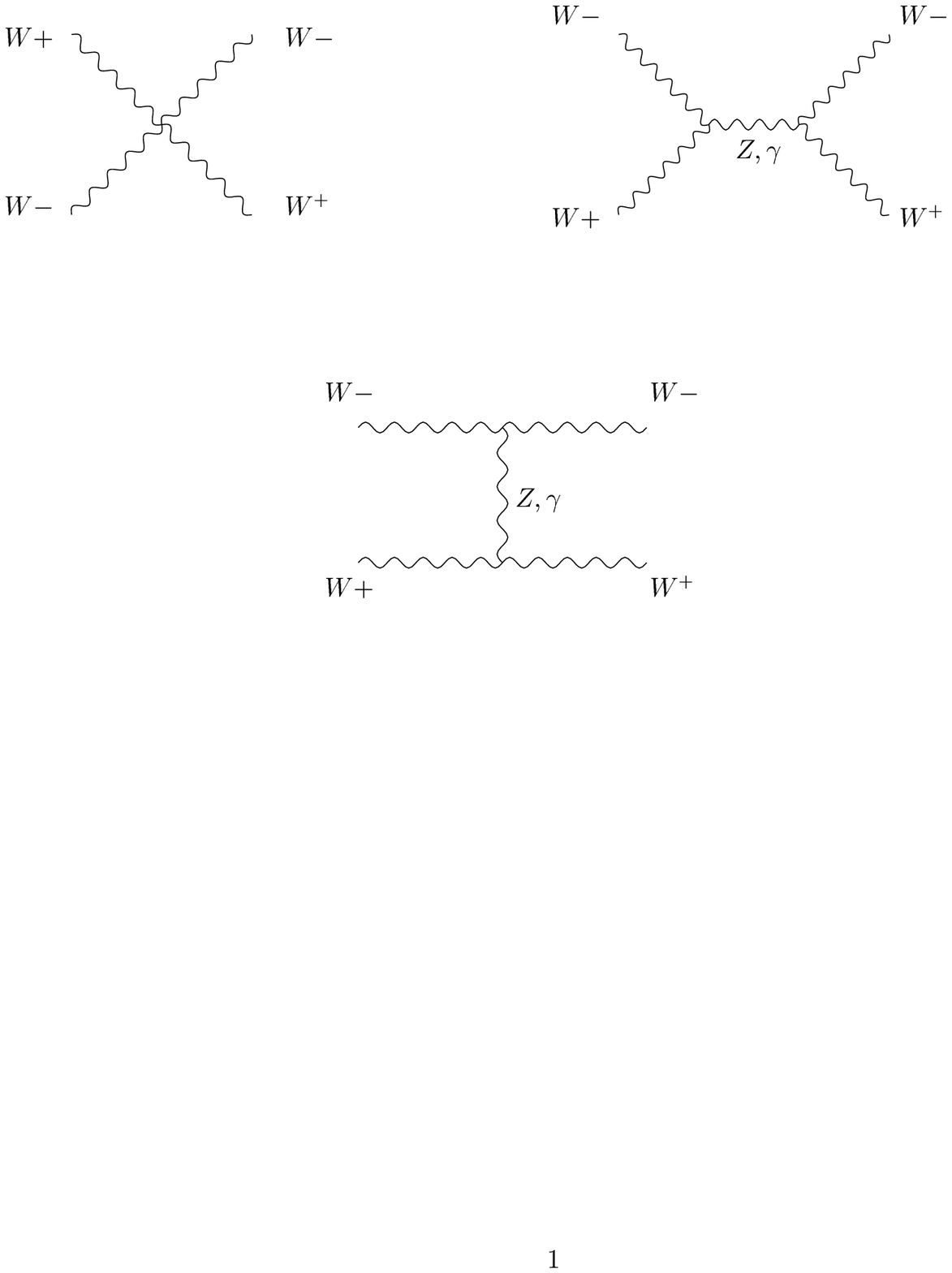}
\vskip -2.2in
\hskip -.2in
\caption{Feynman diagrams contributing to the process $W^+W^-\rightarrow
W^+W^-$ with the Higgs boson removed from the theory.}
\label{fig:unit}
\end{figure}

The second important role of the Standard Model Higgs boson is to
unitarize the gauge boson scattering amplitudes. The $J=0$
partial wave amplitude for the process $W^+W^-\rightarrow
W^+W^-$ (Fig.~ \ref{fig:unit}) grows with energy when the Higgs
boson is not included in the amplitude
and violates partial wave unitarity at
an energy around $E\sim 1.6~TeV$.\cite{unitviol}   The Higgs boson has
just the right couplings to the gauge bosons to restore partial
wave unitarity as long as the Higgs boson mass is less than
around $M_H < 800~GeV$.
With a Higgs boson satisfying this limit, the Standard Model
preserves unitarity at high energies and is weakly interacting.

\section{Problems in Paradise}

The Standard Model  is theoretically unsatisfactory, however,
 because when
loop corrections are included, the Higgs boson mass contains a quadratic
dependence on physics at some unknown higher energy scale, $\Lambda$. 
When the one-loop corrections to the Higgs boson mass, $\delta M_H^2$,
 are computed we find,
\begin{eqnarray}
\delta M_H^2 &=& {G_F\Lambda^2\over 4 \sqrt{2}\pi^2}
\biggl(6 M_W^2 +3 M_Z^2 +M_H^2-12 M_t^2\biggr)
\nonumber \\
&\sim&- \biggl({\Lambda\over .7~TeV}~200~GeV\biggr)^2.
\label{mh2}
\end{eqnarray}
In order to have a light Higgs boson as required by the precision
electroweak measurements, the scale $\Lambda$ must be near $1~TeV$.
The quantum corrections thus  suggest that there must be some new physics
lurking at the $TeV$ scale.

We therefore need new physics at the $1~TeV$ scale to get a light Higgs boson.
However,  much of the possible new physics at this scale is already
 excluded experimentally. A model independent analysis which looked
at various dimension-$6$ operators found that
typically new physics cannot occur below a scale ~$\Lambda>5~TeV$. A 
representative
sampling of limits on possible dimension-6
 operators is shown in Table 1 and a
more complete list can be found in Ref.[3]. 
This tension between needing a low scale $\Lambda$ for new physics
in order to get a light Higgs boson and the experimental exclusion
of much possible new physics at the $TeV$ scale 
has been dubbed the ``little hierarchy problem''.
 However, a global fit to 21 flavor- and CP- conserving
 operators  found that there
are certain directions in parameter space where the limit on $\Lambda$
can be lowered considerably\cite{skib} (even to below
$1~TeV$) raising the possibility that
in specific models the ``little hierarchy problem'' may not be
a problem at all.

\begin{table}%1
\caption{Representative limits ($90~\%$ c.l.) on the scale of new
dimension-6 operators
corresponding to $L={\cal O}_i/\Lambda^2$[4].}
\label{tab:lit}
\begin{tabular}{|c|c|c|} 
 
\hline 
 
\raisebox{0pt}[12pt][6pt]& {}
 
\raisebox{0pt}[12pt][6pt]{Operator, ${\cal O}_i$} & 
 
\raisebox{0pt}[12pt][6pt]{$\Lambda_{min}~(TeV)$ }\\
 
\hline
 
\raisebox{0pt}[12pt][6pt]{LEP} & 
 
\raisebox{0pt}[12pt][6pt]{$H^\dagger\tau HW_{\mu\nu}^a B^{\mu\nu}$} & 
 
\raisebox{0pt}[12pt][6pt]{$10$} \\

\hline

\raisebox{0pt}[12pt][6pt]{LEP-2} & 
 
\raisebox{0pt}[12pt][6pt]{${\overline e}\gamma_\mu e {\overline l}
\gamma^\mu l$} & 
 
\raisebox{0pt}[12pt][6pt]{$5$} \\
\hline
 
\raisebox{0pt}[12pt][6pt]{Flavor} & 
 
\raisebox{0pt}[12pt][6pt]{$H^\dagger {\overline d}_R\sigma_{\mu\nu}
q_L F^{\mu\nu}$} & 
 
\raisebox{0pt}[12pt][6pt]{$9$} \\

\hline
\end{tabular}
\end{table}

In recent years, there have been a variety of creative new models
constructed which
attempt to find a mechanism to lower the scale $\Lambda$, while at the
same time not violating the existing experimental limits.  Supersymmetric
models are the trusty standard for addressing this problem and we
discuss progress and variations on the minimal supersymmetric model in
the next section.  In the following sections, we discuss attempts to
address electroweak symmetry breaking with Little Higgs models\cite{lh,lh1}
 and with
Higgsless models.\cite{higgsless}
  There are many other novel models for electroweak
symmetry breaking--fat Higgs models,\cite{fathiggs} 
 strong electroweak symmetry breaking\cite{sews} (and many more!)
--which will not be
addressed here due to space limitations.

\section{Supersymmetry}

The classic model of new physics at the TeV scale is 
supersymmetry, where a cancellation between the contributions
of the Standard Model particles and the new partner particles
of a supersymmetric model keeps the
Higgs boson mass at the TeV scale. This
cancellation occurs as long as the supersymmetric partner
particles have masses on the order of the weak scale.  For example,
the top quark contribution to Eq.~ \ref{mh2} becomes,\cite{susycan} 
\begin{equation}
\delta M_H^2 \sim  G_F\Lambda^2 \biggl(M_t^2-
{\tilde m}_{t1,t2}^2\biggr),
\label{susymh}
\end{equation}
where ${\tilde m}_{t1,t2}$ are the masses of 
the scalar partners of the top quark.

\begin{figure}%1
\epsfxsize120pt
\figurebox{120pt}{160pt}{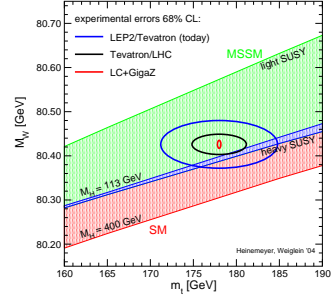}
\caption{Fit to precision electroweak data in the MSSM. The curve
labelled heavy SUSY assumes the supersymmetric parameters are
set at $2~TeV$[11].}
\label{fig:mwmt}
\vskip -.2in
\end{figure}

The simplest version of a supersymmetric model, the MSSM,
 has many positive aspects: 
\begin{itemize}
\item The MSSM predicts gauge coupling unification at the GUT scale. 
\item
The MSSM contains a dark matter candidate, the LSP (Lightest Supersymmetric 
Particle). 
\item
The MSSM  predicts a light Higgs boson, $M_H < 140~GeV$.
\item 
The MSSM agrees with precision electroweak measurements.\cite{susyfit}
\end{itemize}
The fit to the electroweak precision data 
can be performed
in the context of the MSSM and is shown in Fig. \ref{fig:mwmt} for
supersymmetric partner
 masses below $2~TeV$.  The MSSM with supersymmetric
partner particles
in the $1-2~TeV$ region is actually a slightly better statistical 
fit to the data than the Standard Model.\cite{mssmfit} 

There are also many negative things about the supersymmetric
model, the most obvious of which is:  {\bf{{\it Where is it?}}} 

 In the MSSM, the
lightest Higgs boson mass has a theoretical upper bound,  
\begin{eqnarray}
M_H^2 &<& M_Z^2 \cos^2 2 \beta
\nonumber \\
&&
+{3G_F M_t^4\over 
\sqrt{2}\pi^2\sin^2\beta}\log\biggl({
{\tilde m}_{t1}{\tilde m}_{t2}\over M_t^2}\biggr),
\end{eqnarray}
where 
$\tan\beta$ is the ratio of the neutral Higgs boson vacuum expectation
values.
Requiring that the Higgs boson mass
 satisfy the LEP direct search limit, $M_H>114~ GeV$,
implies that the stop squarks must be relatively heavy,\cite{rg}
\begin{equation}
{\tilde m}_{t1}{\tilde m}_{t2}>(950~GeV)^2.
\label{susymass}
\end{equation}
However, the supersymmetric
partner  particles in the MSSM are naturally on the order
of the weak scale, so there
is a tension between the desire for them to  be light (to fill their
required role in cancelling the quadratic contributions to the Higg
boson mass as in Eq.~ \ref{susymh} ) and the limit of Eq.~ \ref{susymass}.

\begin{figure}
\vskip -.5in
\epsfxsize220pt
\hskip -1in
\begin{center} 
\hskip -.25in\figurebox{150pt}{420pt}{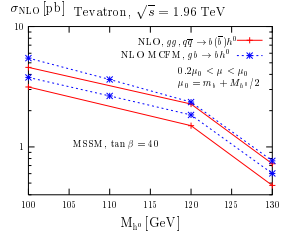}
\end{center}
\vskip -2.in
\caption{Total next-to-leading order cross section in the MSSM for $bH$
production at the Tevatron.  The bands show the renormalization/factorization
dependence. The solid (red) curves correspond to the four-flavor number
scheme with no $b$ partons, and the dotted (blue) curves are
the prediction from  the five-flavor
number scheme with $b$ partons in the initial state[15].}
\label{fig:bbhmssm}
%\vskip -.15in
\end{figure}

\begin{figure}
\epsfxsize160pt
\figurebox{160pt}{200pt}{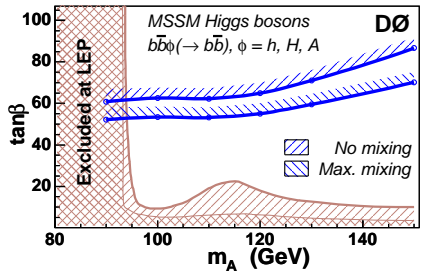}
\caption{$95\%$ c.l. upper limit  from the D0
experiment at the Tevatron on $\tan\beta$ in the MSSM from
$gg\rightarrow b {\overline b} \phi$, where $\phi$ is any of the three
neutral Higgs bosons of the MSSM[17].}
\vskip -.2in
\label{fig:bbh}
\end{figure}

The couplings of the Higgs boson to the bottom quark are enhanced in 
the MSSM for large values of $\tan\beta$ and the dominant
production mechanism becomes $gg\rightarrow b {\overline b} H$, where
$0$, $1$, or $2$ $b$ quarks are tagged.\cite{bbh2,mssmb}
 Fig.~ \ref{fig:bbhmssm} shows
the total next-to-leading order cross section for $bH$ production at
the Tevatron as a function of the mass of the 
lightest Higgs boson of the MSSM for $\tan\beta=40$.\cite{bbh2}
  D0 has a new limit
on this process, which is shown in Fig.\ref{fig:bbh}.\cite{d0bbh} 

Many variants of the MSSM have been constructed.
  One of the simplest is the NMSSM (next-to-minimal-
supersymmetric model)
which is obtained by adding a Higgs singlet superfield ${\hat  S}$ to the
MSSM.\cite{nmssm,nmssm2}  The superpotential in the NMSSM is,
\begin{equation}
W=W_{MSSM}+\lambda {\hat H_1} {\hat H_2}
{\hat S}+{\kappa\over 3}{\hat S}^3,
\end{equation}
where ${\hat H_1}$ and ${\hat H_2}$ are the Higgs doublet
 superfields of the MSSM, and ${\hat S}$ is
the Higgs singlet superfield. 
When the scalar component of
the singlet, $S$, gets a vacuum expectation
value, the term $\lambda {\hat H_1}
{\hat  H_2} <S>$ in the superpotential naturally
generates the $\mu {\hat H_1}
{\hat H_2}$ term of the MSSM superpotential and it
is straightforward to understand why $\mu\sim M_Z$. 
This is the major motivation for constructing the NMSSM.

 In
the NMSSM  model, the bound on the lightest Higgs boson
mass becomes,
\begin{eqnarray}
M_H^2&<&M_Z^2 \cos^2 2 \beta +v^2\lambda^2\sin^2 2 \beta
\nonumber \\
&&  +{\hbox{1-loop~corrections}},
\end{eqnarray}
and the lightest Higgs boson can be significantly
heavier than in the MSSM. 
If we further
assume that the couplings remain perturbative to the GUT scale,
the theoretical upper
bound on the lightest Higgs boson mass becomes $M_H<150~GeV$.\cite{nhiggs}

\begin{figure}
\epsfxsize140pt
\figurebox{140pt}{180pt}{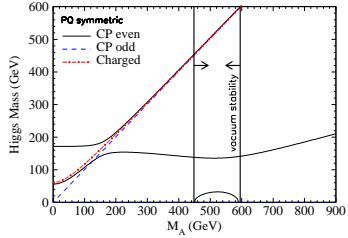}
\caption{Typical mass scenario for the Higgs bosons in the NMSSM
with $\tan\beta=3$. The region between the vertical lines (denoted
by arrows) is the region allowed by vacuum stability[19].}
\label{fig:nmssm}
\vskip -.2in
\end{figure}

The phenomenology in the NMSSM is significantly different
than in the MSSM. There are three neutral
Higgs bosons and two pseudoscalar Higgs bosons.
A typical scenario for the masses is shown in Fig. \ref{fig:nmssm}. 
New decays such as the Higgs pseudoscalar  into 
two scalar Higgs bosons are possible 
and  changes the LHC Higgs search
strategies.
In addition, the lightest Higgs boson can have a 
large CP-odd component and so can evade the LEP bound on 
$M_H$.\cite{nmssm,nmssm2} 

The minimal version of the MSSM  conserves CP, but
CP violation in the Higgs sector can easily be accommodated in the MSSM.
Non-zero phases in the scalar
tri-linear couplings can generate large CP violating
effects from radiative corrections, 
especially those involving the third generation. If there is CP violation 
in the Higgs sector of the MSSM, then the three neutral Higgs mass 
eigenstates, $H_1$, $H_2$, and $H_3$, are mixtures of the CP- even and CP- 
odd Higgs states.\cite{cpx}
 The production and decay properties of the Higgs bosons can 
be very different from those of
the Higgs bosons in the CP conserving version of the MSSM 
since the CP- odd
components of the Higgs mass
eigenstates do not couple to the $Z$ boson. 

Experimental searches for the Higgs boson in a version of the MSSM with 
CP violation in the Higgs sector have been performed by the LEP 
collaborations\cite{igk} using the benchmark parameters of the CPX 
model.\cite{cpx}
 For large values of $M_{H_2}$, $H_1$ is almost completely CP- even 
and the exclusion limit for the lightest Higgs boson mass is similar 
to the CP conserving limit.  If $M_{H_2}>130~ GeV$, then $M_{H_1} > 113~ GeV$. 
 For lighter $M_{H_2}$,  the $H_1$ has a large mixture of the CP- 
odd component and the result is that there are unexcluded regions in the 
$M_{H_1}-\tan\beta$ parameter space and
the excluded region disappears completely for 
$4< \tan \beta <10$. 
At $95\%$~ c.l., $
 \tan\beta < 3.5~{\hbox {and}}~ M_{H_1} < 114~GeV$ and also
$\tan\beta >2.6 $
are excluded in the CPX scenario.\footnote{These limits assume
$M_t=179.3~GeV$[22].}

\begin{figure}
\epsfxsize160pt
\figurebox{160pt}{200pt}{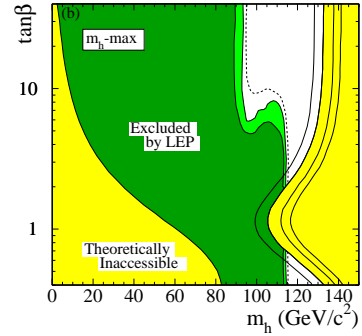}
\caption{Excluded region in the $M_{H_1}$-$\tan\beta$ plane
in the  CP conserving version of the MSSM.
The light (dark) green is the 95~\%  (97 \%cl) exclused region in the
$M_H$(max) benchmark scenario.  The solid lines from left to right
vary the top quark mass: $M_t=169.3,~ 174.3, ~179.3$ and $183~GeV$[21].}
\label{fig:mssmcpyes}
\vskip -.25in
\end{figure}

\begin{figure}
%\vskip -5.in
\epsfxsize160pt
\figurebox{160pt}{200pt}{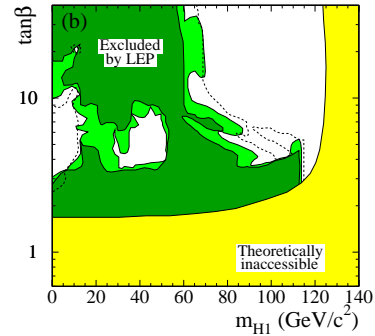}
\caption{Excluded region in the $M_{H_1}-\tan\beta$
plane in the CPX CP violating version of the MSSM. The curves
are as in Fig.~\ref{fig:mssmcpyes}[21].}
\label{fig:mssmcpno}
\end{figure}

  It is interesting to compare the
excluded regions in
the $M_{H_1}-\tan\beta$ plane for
 the CP conserving and CP nonconserving 
versions of the MSSM, as shown in Figs.
\ref{fig:mssmcpyes} and \ref{fig:mssmcpno}.
We observe that the shape of
the excluded region is significantly different in the two cases.
  As noted
in Ref.[22], 
the limit is extremely sensitive to small variations in the top
quark mass.

\section{Little Higgs Models}

\begin{figure}%1
\epsfxsize120pt
\figurebox{120pt}{160pt}{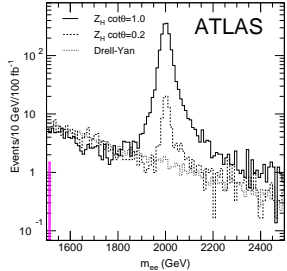}
\caption{ATLAS simulation with $300~fb^{-1}$of data  of the $e^+e^-$invariant mass distribution in a Little Higgs model
resulting from the
decay $Z_H\rightarrow Z H$ for $Z_H=2~TeV$.  The lower dotted histogram
is the background[26].}
\label{fig:lhatlas}
\vskip -.2in
\end{figure}

Little Higgs models\cite{lh,lh1}
 are an attempt to address the  hierarchy problem by cancelling the
quadratic contributions to  the Higgs boson mass
in the Standard Model with the 
contributions resulting from the
addition of new particles which are assumed 
to exist at a scale
around  $1-3~TeV$. The cancellation
of the quadratic contributions occurs between states with the
same spin statistics.  Thus contributions to
Eq.~\ref{mh2} from the Standard Model $W$, $Z$, and photon are
cancelled by the
contributions from new
 heavy gauge bosons, $W_H,Z_H$ and $A_H$, with 
Standard Model quantum numbers, while Standard
Model contributions
from the top quark are cancelled by those from a heavy charge $2/3$ 
top-like quark, and those from the Higgs doublet by 
contributions from a scalar triplet.   
A clear prediction of the Little Higgs models is the existence of these
new particles.  Decays such as $Z_H\rightarrow Z H$ should be particularly
distinctive\cite{lh1} as demonstrated in Fig.\ref{fig:lhatlas}.\cite{atlaslh}

The basic idea of the Little Higgs models is that a continuous global symmetry 
is broken spontaneously and the Higgs boson is the Goldstone boson of
the broken symmetry.  There are many variants of this idea,  with the simplest 
being a model with 
a global SU(5) symmetry broken to a global SO(5) symmetry
 by the vacuum
expectation value of
a non-linear sigma field $\Sigma=exp(2 i \Pi/f)$. The
Goldstone bosons contain both a Higgs doublet
and a Higgs triplet and reside in the field $\Pi$.
  The parameter $f$ sets the scale
of the symmetry breaking, which occurs
at a scale $\Lambda\sim 4 \pi f\sim 10~TeV$ where
the theory becomes strongly interacting.
 The quadratic contributions to the Higgs boson
mass of the Standard Model are cancelled by the
new states at a scale $gf\sim 1-3~TeV$.  
Furthermore, the
gauge symmetries are arranged in such
a manner that the Higgs boson
gets a mass only at two-loops, $M_H\sim g^2 f/(4\pi)$, and 
so the Higgs boson is naturally light, as required by the precision
electroweak data.

\begin{figure}%1
\vskip -.05in
\epsfxsize120pt
\figurebox{120pt}{160pt}{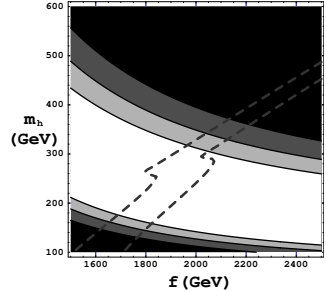}
\caption{Excluded regions at 95\%, 99\% and 99.9\% confidence
level (from lightest to darkest) in the little Higgs model with T-Parity.
In the band between the two dashed lines the lightest T-Parity odd particle
is a consistent dark matter candidate and contributes to a relic
density within $2\sigma$ of the WMAP data[25].}
\label{fig:tparity}
\vskip .4in
\end{figure}

The mixing of the Standard Model gauge bosons with the heavy gauge bosons
of Little Higgs models typically gives strong constraints
on the scale $f>1-4~TeV$.\cite{lhlims}  It is possible to evade many of
these limits by introducing a symmetry ($T$ parity) which requires that
the new particles be produced in pairs.\cite{tparity,tparity2}  
This allows the scale
$f$ to be as low as $500~GeV$. The lightest particle with $T$-odd parity
is stable and is a viable dark matter candidate for $M_H$ between around
$200$ and $400~GeV$ and the scale $f$ in the $1-2~TeV$ region, as seen
in Fig.~\ref{fig:tparity}.  

\begin{figure}%1
%\vskip -1.in
\epsfxsize120pt
\figurebox{120pt}{160pt}{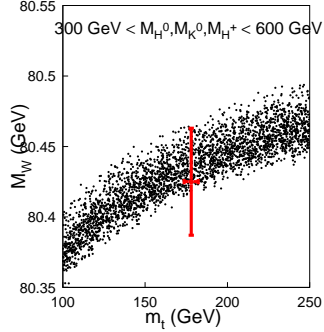}
\caption{Predictions for the $W$ mass as a function of the top quark mass
in a theory with a Higgs triplet.  The masses of the three scalars in
the theory, $H^0,~K^0$, and $H^\pm$, are varied between $300$ and $600~GeV$.
The red point is the experimental data point with the $1\sigma$ errors[27].}
\label{fig:wmass}
\end{figure}

Little Higgs models allow the lightest neutral Higgs boson to be quite heavy,
as is demonstrated in Fig.~\ref{fig:wmass}.\cite{trip}   
 The relaxation of the strong upper bound on the Higgs mass
of the Standard Model is a generic feature of models with
Higgs triplets.

\vskip -.25in

\section{Higgsless Models}

Finally, we consider a class of models in which the Higgs boson is 
completely removed from the theory.  These models face a number of
basic challenges: 
\begin{itemize}
\item
 How to break the electroweak symmetry?  
\item How to 
restore unitarity without a Higgs boson? 
\item How to generate gauge
boson and fermion masses?
\item How to ensure 
\begin{equation}\rho={M_W^2\over
M_Z^2 \cos^2\theta_W}=1?\\ \nonumber\end{equation}
\end{itemize}

\begin{figure}%1
\epsfxsize120pt
\figurebox{120pt}{160pt}{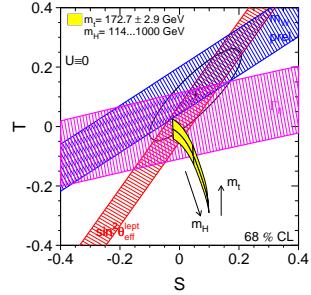}
\caption{Limits on $S$ and $T$ from precision electroweak measurements,
as of September, 2005. The Standard Model reference values (which give 
$S=T=U=0$) are $M_t=175~GeV$ and $M_H=150~GeV$.[1]}
\label{fig:stu}
\end{figure}

Models with extra dimensions
 offer  the possibility of removing the
Higgs boson   from the theory and generating the  
electroweak symmetry breaking from boundary conditions
on the  branes of the extra
dimensions.\cite{higgsless}
 Before even constructing such a Higgsless model, it is
obvious that models of this class will
have problems with the electroweak
precision data.  As can be seen from Fig.\ref{fig:stu}, as the Higgs boson
gets increasingly massive, the predictions of the 
Standard Model get further 
and further away from the data.  
A heavy Higgs boson gives too large a value of $S$ 
and too small a value of $T$.
This figure gives a hint
as to what the solution must eventually be:  
The Higgsless models must have a large and positive
contribution to $T$ and must not have any additional
contributions to $S$.\cite{post}

The Higgsless models  all contain a tower of Kaluza Klein (KK) particles, 
$V_n$, with the quantum numbers of the Standard Model
gauge bosons. 
The lightest particles in the KK tower are the Standard Model $W$, $Z$,
and $\gamma$.
  These Kaluza Klein particles contribute to the elastic scattering
amplitudes for gauge bosons.  In general, the elastic scattering amplitudes
have the form, (where $E$ is the scattering energy):
\begin{equation}
A=A_4{E^4\over M_W^4}+A_2{E^2\over M_W^2}+A_0+...
\end{equation}
In the Standard Model, $A_4$ vanishes by gauge invariance and $A_2$ vanishes
because of the cancellation between the gauge boson and Higgs boson
 contributions.
In the Higgsless models, the  contributions to $A_4$ and $A_2$ cancel if,
\begin{eqnarray}
g^2_{nnnn}&=&\Sigma_kg^2_{nnk}
\nonumber\\
4g^2_{nnnn}&=&3\Sigma_kg^2_{nnk}{M_k^2\over M_n^2 },
\end{eqnarray}
where $g_{nnk}$ is the cubic coupling between $V_n$, $V_n$, and $V_k$,
$g_{nnnn}$ is the quartic self coupling of $V_n$, and $M_k$ is the mass
of the $k^{th}$ KK particle.

 The amazing
fact is that the 5-dimensional Higgsless models satisfy these sum rules
exactly due to 5-dimensional gauge invariance.  Similarly, 4-dimensional
deconstructed versions of the  Higgsless models\cite{deconstruct}
 satisfy these sum rules to an accuracy
of a few percent. The Kaluza Klein particles play the same role as the
Higgs boson does in the Standard Model and unitarize the scattering amplitudes.
Of course, the lightest Kaluza Klein mode needs to be light enough for 
the cancellation to occur before the amplitude is already large, which
restricts the masses of the Kaluza
Klein particles to be less than $1-2~TeV$.\cite{kkunit,kkunit2}

\begin{figure}%1
\epsfxsize120pt
\figurebox{120pt}{160pt}{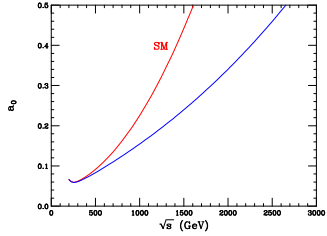}
\caption{$J=0$ partial wave for elastic gauge boson scattering
in the Standard Model with the Higgs boson removed (red)  and with the 
inclusion of a single Kaluza Klein excitation with $M=500~GeV$
(blue) in a deconstructed Higgsless model[30].}
\label{fig:higgsj0}
\end{figure}

Fig.~ \ref{fig:higgsj0} shows the growth of the $J=0$ partial wave in the
Standard Model with the Higgs boson removed and in a Higgsless model
with a single Kaluza Klein particle with mass $M=500~GeV$ included. The
inclusion of the Kaluza Klein contributions pushes the scale of unitarity
violation from $E\sim 1.6~GeV$ in the Standard Model with no Higgs boson to 
around $E\sim 2.6~TeV$ in the Higgsless models. 

The Kaluza Klein particles contribute to the electroweak precision 
measurements. In general, the corrections are too large for KK particles
with masses on the $TeV$ scale.\cite{kkstu}
Considerable progress in addressing this problem
has been made in the last year with the realization that the contributions
of the Kaluza Klein particles to the precision 
electroweak observables depend
on where the 
fermions
 are located in the extra dimensions.  In the Randall-Sundrum
model, $S$ is positive if the fermions are located on the Planck brane and
negative if they are located on the $TeV$ brane.  The trick is to find
an intermediate point where there is a weak coupling between 
the KK modes and the fermions.\cite{kkstu,kkstu2}  It appears to be possible
to construct models which are consistent with the electroweak precision
measurements by having the fermion wavefunction be located
between the branes.\cite{kkstu2}

\begin{figure}
%\vskip -.5in
\epsfxsize200pt
\figurebox{200pt}{2400pt}{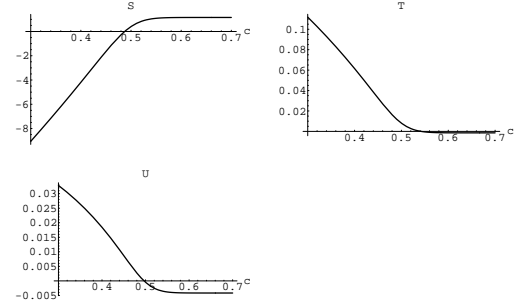}
\caption{Oblique parameters, $S$, $T$, and 
$U$ in a Higgsless model as a function of the
fermion localization parameter, $c$. If the fermions
 are localized on the $TeV$ brane, $c << {1\over 2}$, 
while fermions localized on the Planck brane have $c>> {1\over 2}$.  A flat
fermion wavefunction corresponds to $c={1\over 2}$[32].}
\label{fig:stuhiggs}
\vskip -.25in
\end{figure}

  Fig.~ \ref{fig:stuhiggs} shows the oblique parameters as a function
of the variable $c$, which characterizes the location of the fermion
wavefunction.  If the fermions are localized on the $TeV$ brane, 
$c << {1\over 2}$, while fermions localized on the Planck brane have 
$c>> {1\over 2}$.  A flat
fermion wavefunction corresponds to $c={1\over 2}$.  
For $c\sim {1/2}$ it is
possible to satisfy the bounds from precision electroweak data.  Fermions with a flat wavefunction are weakly coupled to the Kaluza Klein
particles and so such Kaluza Klein particles would have escaped the
direct searches for heavy resonances at the Tevatron.

The next challenge for Higgsless models
 is to generate the large mass
splitting between the top and the bottom quarks.\cite{tb}

\begin{figure}%1
\vskip -.1in
\epsfxsize120pt
\figurebox{120pt}{160pt}{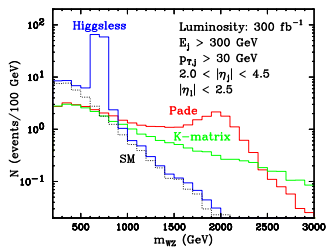}
\caption{The number of events per $100~GeV$ bin in the 2-jet plus 3 lepton
plus $\nu$ channel at the LHC, coming from the subprocess
$WZ\rightarrow WZ$ in a Higgsless model[34].}
\label{fig:lhc}
\end{figure}

Weakly coupled Kaluza Klein particles are a generic feature of 
Higgsless models and can be searched
for in a model independent fashion.  These KK particles appear as
 massive $W$-,
$Z$-, and $\gamma$- like resonances in vector boson fusion and they
will appear as narrow resonances in the $WZ$ channel as shown in
Fig.~ \ref{fig:lhc}.\cite{bmp}  The lightest KK
resonance should be clearly
observable above the background.

\section{Conclusions}
The mechanism of electroweak symmetry breaking could be far more
complicated than a simple Higgs boson.  Almost all models, however,
have distinctive signatures which should be observed at the LHC.
Soon, with data from the LHC,
 we should have some indication what mechanism nature has
chosen!  A complete understanding of the unknown physics awaiting
us at the TeV scale will probably require a future linear collider.\cite{lc}

\section*{Acknowledgments}
This research supported by Contract No. DE-AC02-76CH1-886 with the
U.S. Department of Energy. I thank my collaborators, M.C. Chen, C.~
Jackson, T. ~Krupovnickas,  L.~Reina, and D.~Wackeroth for countless
discussions.

%%%%%%%%%%%%%%%%%%%%%%%%%%%%%%%%
%  Question and Answer Section %
%%%%%%%%%%%%%%%%%%%%%%%%%%%%%%%%
% Use clear page to make sure everything is flush and a new
% page is started (not just a new column)
%%%%%%%%%%%%%%%%%%%%%%%%%%%%%%%%
%\clearpage
\section*{DISCUSSION}

\begin{description}
\item[Daniel Kaplan] (Illinois Institute of Technology):
\\
How does the new state possibly seen in the HyperCP experiment at Fermilab fit
into SUSY models? It has a mass of 214.3 MeV and decays into $\mu^+ \mu^-$.

\item[Sally Dawson{\rm :}]
 
This state is very difficult to understand in terms of SUSY models. 
 
\item[Anna Lipniacka] (University of Bergen):

Is gauge coupling unification natural in Large Extra Dimension models?

\item[Sally Dawson{\rm :}]

No.  These theories typically violate unitarity and become strongly interacting
at a scale between $1$ and $10$~ TeV.

\item[Ignatios Antoniadis] (CERN):

What is the prize to pay in models that solve the little hierarchy problem,
such as the little Higgs models, in particular on the number of parameters and
the unification of gauge couplings?

\item[Sally Dawson{\rm :}]
Obviously, there is a large increase in the number of parameters and gauge
unification is forfeited. 

\item[Luca Silvestrini] (Munich and Rome):\\
Maybe one should comment about the statement that you made that new physics has
to have a scale $\Lambda$ greater than 5 TeV. Of course this is a conventional
scale that is only valid if the coupling in front of the operator is one, which
is generally not true in any weakly interacting theory and generally not true
if new physics enters through loops. So I do not want that anybody in the
audience really believes that new physics must be at a scale larger than 5 TeV.
It can easily be around the electroweak scale as we know very well.

\item[Sally Dawson{\rm :}]
Absolutely true.  The limits depend on the couplings to the operators, which in
turn depend on the model.

\end{description}
 
\end{document}